\documentstyle[aps,prb,psfig,eqsecnum]{revtex}
\begin{document}
\title{First-Principles Based Matrix-Green's Function Approach to 
Molecular Electronic Devices: General Formalism}
\author{ Yongqiang Xue \thanks{Author to whom correspondence should be 
addressed. Present address: Department of Chemistry and Materials Research Center, 
Northwestern University, Evanston, IL 60208. Electronic mail: ayxue@chem.nwu.edu} }
\address{
Department of Chemistry and Materials Research Center, Northwestern 
University, Evanston, IL 60208 
and School of Electrical and Computer Engineering, Purdue University, 
West Lafayette, IN 47907}
\author{Supriyo Datta }
\address{
School of Electrical and Computer Engineering, Purdue University, 
West Lafayette, IN 47907}
\author{Mark A. Ratner}
\address{
Department of Chemistry and Materials Research Center, Northwestern 
University, Evanston, IL 60208}

\date{\today}
\maketitle

\begin{abstract}
Transport in molecular electronic devices is different from 
that in semiconductor mesoscopic devices in two important aspects: (1) 
the effect of the electronic structure and (2) the effect of the interface 
to the external contact. A rigorous treatment of molecular electronic 
devices will require the inclusion of these effects in the context 
of an open system exchanging particle and energy with the external 
environment. This calls for combining the theory of quantum transport 
with the theory of electronic structure starting from the 
first-principles. We present a rigorous yet tractable matrix Green's 
function approach for studying transport in molecular electronic 
devices, based on the Non-Equilibrium Green's Function Formalism of quantum 
transport and the 
density-functional theory of electronic structure using local orbital 
basis sets. By separating the device rigorously into the molecular 
region and the contact region, we can take full advantage of the 
natural spatial locality associated with the metallic screening in the 
electrodes and focus on the physical processes in the finite molecular 
region. This not only opens up the possibility of using 
the existing well-established technique of molecular electronic structure 
theory in transport calculations with little change, 
but also allows us to use the language of qualitative molecular orbital 
theory to interpret and rationalize the results of the computation. 
We emphasize the importance of the self-consistent charge transfer 
and voltage drop on the transport characteristics and describe the 
self-consistent formulation for both device at equilibrium and device 
out of equilibrium. For the device at equilibrium, our method provides an 
alternative approach for solving the molecular chemisorption problem. 
For the device out of equilibrium, we show that the calculation of 
elastic current transport through molecules, both conceptually 
and computationally, is no more difficult than solving the chemisorption 
problem. 
\end{abstract}

\section{Introduction}
\label{S1}

There has been significant progress in exploring 
the concept of molecular electronics in recent years, due to the advancement 
of techniques for characterizing and manipulating individual 
molecules~\cite{Feynman,Ratner74,Ratner98,ReedR99,DekkerR99,Joachim00}. The 
fact that useful devices can be built on the basis of individual molecules, 
as demonstrated recently by several research groups
~\cite{Reed97,Reed99,JG98,GJ99,Dekker97,Dekker00,Avouris89,RR95,Datta97,Xue991,Metzger97,Heath99}, has generated wide-spread interest 
in this new technology. In order to fulfil the true promise of molecular 
electronics, it is essential to have a thorough understanding of 
the electronic and transport processes at the single molecule level. This 
paper represents an attempt to put our understanding of electronic 
transport through individual molecules on a firm theoretical basis 
starting from the first-principles.     

Traditionally electron transport phenomena are studied in the context 
of bulk semiconductor devices, the theoretical description of which is 
largely built upon two premises: (1) the effective-mass equation and (2) 
the Boltzmann Transport Equation (BTE)~\cite{Lundstrom}. 
The effective-mass equation 
subsumes the effect of the background periodic lattice potential into an 
effective Hamiltonian so that the electrons can be considered as particles of 
effective mass $m^{*}$ in some applied field. The semi-classical nature of 
the electronic motion in the bulk devices, on the other hand, allows us 
to describe the distribution of electrons in response to 
the applied fields and various scattering mechanisms through the solution 
of the BTE equation, in much the same way as that of classical point-like 
particles. Within this approach, the quantum-mechanical effect only 
comes in through the calculation of band structures and the various 
carrier scattering rates, which provides the input to the solution of 
the BTE or its approximate versions such as the drift-diffusion 
equation. As a result, the study of transport phenomena is effectively 
decoupled from the study of the electronic structures.  

The investigation of quantum mechanical transport has begun to flourish 
during the past two decades, largely due to the 
advancement in lithographic techniques which has allowed routine 
fabrication of submicron features in artificially tailored semiconductor 
heterostructures~\cite{Datta95,Ferry97}. Two quantum mechanical effects 
distinguish such mesoscopic devices from bulk devices, reflecting 
the wave-particle duality of electron. One is the quantization 
of electronic charge which evidences itself in coulomb blockade and 
single-electron transistors~\cite{GD92}. 
The other is the preservation of quantum phase coherence over a length 
with size comparable to one of the device dimensions and the resulting 
energy quantization of confined electrons, which leads to the observation 
of conductance quantization in transport through a narrow constriction 
(quantum point contact) and negative differential resistance in 
double-barrier tunneling structures (resonant-tunneling diode). 
Transport in these quantum semiconductor structures is now determined 
by the scattering property and occupation of the electronic eigenstates 
under an appropriate external potential. On the other hand, electrons in 
such quantum semiconductor structures are usually confined in one or 
two directions. The confinement potential, often the electrostatic 
potential due to a nearby gate or the band discontinuity across the 
heterostructure interface, varies slowly on the atomic scale. So the 
electrons can still 
be described satisfactorily as free carriers using the single-band 
effective mass approximation and most of the important phenomena in 
mesoscopic transport can be understood without worrying about the complex 
electronic structures~\cite{Datta95,Ferry97}.    

The situation changes dramatically when we consider molecular electronic 
devices formed by sandwiching a chemically synthesized molecule between 
two large (on the molecular scale) metallic electrodes. At the molecular 
scale, the simplicity associated with the effective-mass approximation 
breaks down and the electronic structure of the system has to be taken 
explicitly into account. The quantum mechanical scattering problem involved  
is now the scattering of electrons under the potential of the atomic nuclei 
\emph{and} the potential due to other electrons. Understanding transport 
in such molecular devices therefore requires a knowledge of the microscopic 
electronic structure of the electrode-molecule-electrode system under both 
zero and finite bias voltages. 

The reduction to molecular scale also brings in another complication in 
transport study as compared to the mesoscopic systems: treatment of the 
interface to an external contact. In mesoscopic transport, the details of 
contact are often not important. The measuring electrodes, taken as 
infinite electron reservoirs, can either be simulated by reflectionless 
semi-infinite leads with simple confinement potential at the interface, or 
only come into the theoretical formulation as an appropriate boundary 
condition~\cite{Datta95,Ferry97}. This is no longer true when the 
device is of molecular dimension. Since the electrodes can have atomic 
structures on the surface 
whose dimensions can be comparable to the molecule, the usually 
well-defined boundary between the active device region and the 
contact region is blurred. The interface to the external contact 
becomes an integral part of the device and the measured electrical 
characteristics will depend on the details of the atomic arrangement of 
the contact. Moreover, the electronic and structural properties of the 
molecule could be modified by the bonding to the measuring contact, 
bringing in additional complications~\cite{Lang78,Hoffmann88,BT99}. 

In summary, molecular electronic devices are different from their 
mesoscopic counterparts in two important aspects: (1) the effect of the 
electronic structure and (2) the effect of the interface to the external 
contact. Since the molecule can freely exchange energy and electrons 
with the electrodes, a rigorous treatment of molecular electronic device 
can only be achieved including these effects in the context of an 
open system. As a result, a successful modeling of molecular electronic 
devices in general calls for combining the theory of quantum transport 
and the theory of electronic structure starting from first-principles. 
  
There have been numerous theoretical works on transport through individual 
molecules describing the electronic structure at different levels. 
Early works have focused on understanding the fundamental mechanisms of 
transport in the molecular-scale and developing simple theory for the 
explanation of experimental results
~\cite{Datta97,Xue991,Ratner94,Ratner99,Datta98,Joachim96,Joachim97,EK99,Hush00,Xue992}. 
These works thus were centered on model Hamiltonian or semi-empirical 
theories (notably the Extended-H\"{u}ckel-Theory of organic 
molecules~\cite{Hoffmann63} and $\pi$-electron tight-binding theory of 
carbon nanotubes~\cite{SDDBook}). 
Due to the interdisciplinary nature of molecular electronics, different 
methods have been used reflecting the authors' own background which are 
essentially all equivalent to the Landauer formula of mesoscopic 
transport~\cite{Landauer70,Buttiker85,Buttiker86} as used extensively 
in our previous works~\cite{Datta97,Xue991,Ratner94,Ratner99,Datta98}. 
Such works have provided useful insights into the factors 
governing transport through individual molecules. However, 
the usefulness of this approach is limited by its incapability of providing 
an accurate description of the electronic structure of the molecule and 
the metal surface involved. Even if a good parameterization exists for 
the molecule and the bulk separately, the charge transfer and the 
resulting self-consistent charge and potential relaxation upon the 
formation of surface and adsorption of the molecule is difficult to 
treat unless drastic assumptions are made or additional parameters are 
introduced. In addition, it is difficult to take into account coupling 
between electrons and molecular vibrational modes which is expected to 
play an increasingly important role as the molecular size 
increases~\cite{Ness01,PH01,Nitzan01}. 

More recently, Lang and coworkers~\cite{Lang95,Lang001,Lang002,Lang003,Lang01} 
have presented self-consistent studies 
of both the conductance and current-voltage characteristics of atomic and 
molecular wires using the jellium model of a metal surface and the 
local-density-approximation (LDA) of density functional theory 
(DFT) with plane-wave basis set.  
The calculation proceeds by recasting the 
Kohn-Sham equation of the electrode-molecule-electrode system into a 
scattering form using the Lippmann-Schwinger equation and solving the 
wavefunctions of the scattering state self-consistently~\cite{Lang95}. 
The current is obtained by summing over the contribution of the 
scattering states which are occupied according to the Fermi distribution 
in each electrode, following the spirit of Landauer theory. Similar 
approaches have also been used by Tsukada and coworkers~\cite{Tsukada95} 
and Guo and coworkers~\cite{Guo97}.  

The jellium model has an appealing and physically reasonable simplicity: 
all the complexities of the bulk band structure are simply ignored, the 
effect of the substrate persists only in providing a continuous energy 
spectrum and the only inhomogeneity left in the metal region is the 
essential one-dimensional inhomogeneity of the surface itself. However, 
the jellium model is known to be deficient in describing the electronic 
density of states and charge density in the region perturbed by the 
absorbed molecule even for sp-bonded simple metals for which it is more 
successful~\cite{BT99,Scheffler94}, and it cannot answer 
questions regarding the effect of adsorption geometry and surface 
relaxation. In addition, it is not applicable to semiconductors and gives 
serious quantitative error when applied to noble and transition metals 
where the bonding at the surface is more directional. The use of the 
plane-wave basis set and the wavefunction formulation also make it 
difficult to treat larger molecular systems, to improve the description 
of the electronic structure and to include other scattering mechanisms. 

The objective of this paper is to present a rigorous yet tractable 
self-consistent matrix Green's function (MGF) approach for studying 
transport in molecular electronic devices, through which many 
limitations of the previous works can be overcome rather conveniently 
using existing computational techniques. In particular, this allows us 
to use existing quantum chemical methods for transport calculation with 
minimum change. The method is based on the Non-equilibrium 
Green's Function Formalism of quantum transport~\cite{HJBook} and the 
density-functional-theory of electronic structure~\cite{DGBook}. 
We also use the technique of expansion in a finite set of local orbital 
functions. 
     
The Non-Equilibrium Green's Function Formalism has proved to be a 
powerful and formally rigorous approach to studying quantum transport 
phenomena in nanodevices, where the devices can be characterized as 
a small active device region connected to electron reservoirs via 
non-interacting leads. Arbitrary interactions in the device can be 
included~\cite{HJBook,MW92}. For non-interacting electron 
systems in the coherent limit, it reduces to the familiar Landauer theory 
of mesoscopic transport~\cite{HJBook,MW92}. 
The formalism is fairly complete, and it has been applied 
successfully to studying quantum-mechanical transport phenomena in 
semiconductor mesoscopic systems such as resonant-tunneling diode and 
single-electron transistors~\cite{Datta95,Ferry97,HJBook}, 
usually in combination with a model Hamiltonian 
and more recently using semi-empirical tight-binding 
formulations~\cite{Lake97}. For molecular electronic devices, within the 
Born-Oppenheimer approximation, the problem is to  
calculate the Green's function of the interacting electron system 
under the potential of the given configuration of atomic nuclei and 
external fields. 

The study of molecular electronic devices is greatly 
facilitated by recognizing the fact that due to the metallic screening in 
the electrodes, the charge and potential perturbation induced by the 
adsorption of the molecule extends over only a finite region into the 
electrodes, practically the region enclosing the surface metal 
atoms closest to the molecule. The charge and potential distributions 
beyond this region are the same as that of the bimetallic interfaces 
without the molecules. By expanding the wavefunction in terms of the 
finite set of local atomic orbital functions, the matrix Green's 
function approach allows us to take full advantage of this spatial 
locality by separating the device into the ``extended-molecule'' 
(which includes the molecule itself and the surface atoms perturbed by 
the molecule) and the electrode region, the description of each can be 
treated and systematically improved independent of each other. The 
effect of the external contact can be included rigorously as a 
self-energy operator, which depends only on the charge and potential 
distribution outside the ``extended molecule'' and needs 
to be computed only once, thereby the computation can be focused on the 
physical processes happening in the finite ``extended molecule'' region.  
This not only opens up the possibility of using the full 
repertoire of molecular electronic structure calculation for studying 
transport in molecular electronic devices, but also allows us to 
interpret the result of computation using the simple picture of bonding 
and orbital interactions in molecules, which is necessary for the 
computation to be useful in terms of understanding rather than pure 
numbers. Although we will focus on the density-functional theory of the 
interacting electron system, the formalism doesn't depend on the 
particular description of the electronic structure.  
  
In matrix form, the present formulation is equivalent to previous works 
using semi-empirical tight-binding formulations~\cite{Datta98,Lake97}. 
The tight-binding formulation provides a conceptually and computationally 
simple framework for studying transport in systems where a quantum 
mechanical description of the electronic structure is necessary but the 
characteristic length scale is much larger than inter-atomic spacing so 
atomic-scale details are not needed. Example systems include layered 
semiconductor devices and long carbon nanotubes. The present formulation 
keeps the simplicity of the tight-binding approach while putting it on 
a firm theoretical basis. The explicit use of the local basis function in 
the real space also allows much more detailed understanding of the physical 
processes through quantities such as the spatial distribution of charge and 
current densities, which play the fundamental role in density-functional 
theory and its time-dependent extension~\cite{DGBook,PYBook,Gross96}. 
In addition, the ambiguities 
associated with the semi-empirical formulation, such as the orthogonality of 
the basis functions used in such formulation and the treatment 
of self-consistent charge-transfer effect~\cite{TB}, can be avoided. 
When the potentials and the integrals are calculated approximately rather 
than from first principles, it can also serve as the basis of improved 
self-consistent tight-binding approaches capable of handling much larger 
systems~\cite{SCCTB}. 

The remainder of this paper is organized as follows: We discuss briefly 
the use of Non-Equilibrium Green's Function and density-functional theory 
for modeling transport in molecular devices in Sec.\ II. 
The matrix Green's function method is described in Sec.\ III.
The self-consistent formulation for device at equilibrium is given in 
Sec.\ IV, where we show that our method provides an alternative and 
generalizing approach to the familiar chemisorption problem in surface 
physics. The self-consistent formulation for a device out of equilibrium 
is given in Sec.\ V. Finally, we devote Sec.\ VI to conclusions. We use 
atomic-units throughout this paper unless otherwise noted.        

\section{Non-Equilibrium Green's Function Approach for Modeling Molecular 
Electronic Devices}

\subsection{Non-Equilibrium Green's Function Formalism}
\label{S2}
As current flows, the device is driven out of equilibrium. For systems 
out of equilibrium, the Green's function approach can be developed 
following the same procedure as the equilibrium case, by defining the 
contour-ordered Green's function~\cite{Keldysh65,Langreth76,Daniel84} 
$G(1,1^{'})=-i\langle T_{C}[\psi_{H}(1)\psi^{+}(1^{'})] \rangle$ (for 
details, see Haug and Jauho \cite{HJBook}). The Green's functions 
involved, besides the retarded and advanced Green's function, 
\begin{eqnarray}
\label{GRGA}
G^{R}(1,1^{'}) &=& -i\Theta (t_{1}-t_{1^{'}})
\langle \{ \psi_{H}(1),\psi_{H}^{+}(1^{'}) \} \rangle , \nonumber \\
G^{A}(1,1^{'}) &=& i\Theta(t_{1^{'}}-t_{1})
\langle \{ \psi_{H}(1),\psi_{H}^{+}(1^{'}) \} \rangle,
\end{eqnarray} 
include the correlation function (or the ``lesser'' Green's function) which 
is the central quantity in this formalism
\begin{equation}  
\label{G<}
G^{<}(\vec r,t;\vec r',t')=
+i \langle \psi^{+}({\vec r,t}) \psi( \vec r',t') \rangle 
\end{equation}
Any physical observable can be obtained from $G^{<}(\vec r,t;\vec r',t')$ 
and its transformations. For example, the quantities of most interest 
to us, the charge density $n(\vec r,t)$ and the current 
density $j(\vec r,t)$, are determined as following:
\begin{equation}
\label{QDens}
n(\vec r,t)= \langle \psi^{+}({\vec r,t}) \psi( \vec r,t) \rangle
           =-i G^{<}(\vec r,t;\vec r,t)
\end{equation}
and
\begin{equation}
\label{IDens}
j(\vec r,t)= \frac{1}{2i} \lim_{\vec r' \to \vec r} (\nabla - \nabla^{'})
             \langle \psi^{+}({\vec r,t}) \psi( \vec r',t') \rangle
           = \frac{1}{2} \lim_{\vec r' \to \vec r} (\nabla^{'} - \nabla) 
             G^{<}(\vec r,t;\vec r',t')
\end{equation}

For steady state, which we consider here, the Green's functions depend only 
on the time difference $t-t^{'}$, which we can Fourier transform to energy. 
The resulting equations of motion (EOM) of the Non-Equilibrium Green's 
function are the Keldysh-Kadanoff-Baym 
equation~\cite{KBBook,Keldysh65,Langreth76}:
\begin{equation}
\label{GKB1}
\{ E-[-\frac{1}{2}\nabla^{2}+V_{ext}(\vec r)] \} G^{R}(\vec r,\vec r';E)
-\int d\vec r''\Sigma^{R}(\vec r,\vec r'';E)
G^{R}(\vec r'',\vec r';E)
=  \delta(\vec r - \vec r'), 
\end{equation}
and
\begin{eqnarray}
\label{GKB2}
\{ E-[-\frac{1}{2}\nabla^{2}+V_{ext}(\vec r)] \} G^{<}(\vec r,\vec r';E)
 &-& \int d\vec r''\Sigma^{R}(\vec r,\vec r'';E)
G^{<}(\vec r'',\vec r';E) \nonumber \\
&=& 
\int d\vec r'' \Sigma^{<}(\vec r,\vec r'';E)G^{A}(\vec r'',\vec r';E)
\end{eqnarray}
or its equivalent integral formulation:
\begin{equation}
\label{Keldysh}
G^{<}(\vec r,\vec r';E)=\int d\vec r'' \int d\vec r'''
G^{R}(\vec r,\vec r'';E)
\Sigma^{<}(\vec r'',\vec r''';E)G^{A}(\vec r''',\vec r';E)
\end{equation} 
where we assume that the non-equilibrium term in the Hamiltonian is 
incorporated into a one-body external potential $V_{ext}$. The interactions 
are contained in the self-energy operators $\Sigma[G]$ which can be obtained 
systematically from many-body perturbation theory following the same 
procedure as their equilibrium counterpart~\cite{Daniel84}.
   
By solving the KKB equation for $G^{<}$, we will be able to calculate the 
current distribution within the device given the Hamiltonian of the system. 
However, in general, they involve the evaluation of the full 
correlation function and the retarded Green's function 
\emph{in the presence of tunneling into the leads}. The usefulness of 
the above expressions then depends on whether we can devise practical 
calculation schemes for the single-particle Green's functions. For 
molecular-scale devices, 
the problem is to determine the state of the interacting electron system 
under the potential of the given atomic nuclei. By separating the term 
describing the electron interactions into the 
classical Coulomb part and the exchange-correlation part, we can write down 
the Keldysh-Kadanoff-Baym equation in the following form: 
\begin{eqnarray}
\label{KKB}
\{ E-[-\frac{1}{2}\nabla^{2}+V_{ext}(\vec r)+\int d\vec r' 
\frac{\rho (\vec{r}')}{|\vec{r}-\vec{r}'|}] \} G^{R}(\vec r,\vec r';E) 
\nonumber \\
-\int d\vec r^{''}\Sigma^{R}(\vec r,\vec r^{''};E)
G^{R}(\vec r^{''},\vec r^{'};E)
= \delta(\vec r - \vec r^{'}), \nonumber \\
G^{<}(\vec r,\vec r^{'};E)=\int d\vec r^{''} \int d\vec r^{'''}
G^{R}(\vec r,\vec r^{''};E)
\Sigma^{<}(\vec r^{''},\vec r^{'''};E)G^{A}(\vec r^{'''},\vec r^{'};E)
\end{eqnarray}
where the external potential $V_{ext}$ is the summation of the ionic 
potential and the potential that drives the system out of equilibrium (see 
the discussion in sections \ref{S4} and \ref{S5} later).
 
\subsection{Density-Functional Theory as the Compromise}

For transport through molecular scale devices, the system is characterized 
by the highly non-equilibrium distribution in the device region, which 
is driven by contact to two large reserviors with different electrochemical 
potential. A truly first-principles treatment of the electronic processes 
in such non-equilibrium system can only be based on the non-equilibrium 
version of the many-body perturbation theory such as the quasi-particle 
theory~\cite{Wilk00} or perhaps the time-dependent density-functional 
theory~\cite{Gross96}. Such 
a first-principles theory doesn't exist yet. Instead, we will work with 
density-functional theory~\cite{DGBook,PYBook}.  
    
The theory of DFT is based on the study of systems in their ground state 
or in thermodynamic equilibrium corresponding to the grand canonical 
ensemble, where the variational formulation of quantum mechanics 
can be used. For transport through non-interacting systems, the problem 
can be solved by working in the scattering state representation, since 
electrons coming from different electrodes are in separate thermodynamic 
equilibrium at different electrochemical potentials. This is the essence 
of the Landauer theory of quantum transport. In this case, it may be 
justified to use density-functional theory to calculate the scattering 
state wavefunctions which correspond to a two-component system, each 
characterized by a well-defined thermodynamic ensemble. However, 
we are not aware of any formal proof of this. In fact, this has been the 
basis for the use of DFT to calculate current and conductance in the 
past~\cite{Lang95,Lang001,Lang002,Lang003,Tsukada95,Guo97}. For the 
interacting electron system which DFT attempts to describe, the nonlinear 
current-voltage characteristic can be obtained only by doing 
perturbation theory in the part of the Hamiltonian that drives the system 
out of equilibrium. For these reasons, the use of the DFT formalism, with 
probably the exception of its time-dependent extension~\cite{Gross96}, in 
transport calculations can only be taken as \emph{qualitative in 
principle}, although this doesn't preclude its \emph{quantitative success 
in practice}.   

From here on, we will work with density functional theory in our modeling 
of molecular devices, with the understanding that we approximate the true 
self-energy operator for exchange-correlation interaction with the DFT 
description of exchange-correlation potential~\cite{Wilk00}. 
We will treat the usage 
of DFT in a more qualitative sense in that it has a well-defined physical 
basis, from which we know where it can be expected to succeed and where 
it may fail and why. Therefore we will put more emphasis on using the 
results of such calculation for qualitative understanding whenever possible. 
These results may serve as the basis for further improvement as our 
understanding of molecular devices progresses. The reason for this 
choice is therefore \emph{practical, rather than fundamental}.

\section{The Method of Matrix Green's Function}

Our starting point is the Keldysh-Kadanoff-Baym equation (Eq.\ (\ref{KKB})).  
Remember that we approximate the self-energy operator $\Sigma$ by the DFT 
description of the exchange-correlation potential $V_{xc}(\vec r,\vec r')$ 
which is an energy independent real operator, but doesn't need to 
be local. In particular, we will only assume $V_{xc}$ to be local in the 
electrode region. It can take any non-local form in 
the ``extended molecule''. There is 
no ``lesser'' self-energy operator $\Sigma^{<}$ associated with $V_{xc}$. 
This is in contrast with the true quasi-particle 
theory where $\Sigma$ is in general non-Hermitian and the corresponding 
$\Sigma^{<}$ represents the scattering rates or ``life-time'' of the 
quasi-particle state~\cite{Daniel84,Hedin69}. 

Given the exchange-correlation potential $V_{xc}(\vec r,\vec r')$ , we can 
define the single-particle wavefunction $\psi_{\mu}(\vec r)$:
\begin{equation}
\label{KS}
[-1/2\nabla^{2}+V_{ext}(\vec r)
+\int\frac{n(\vec r')}{|\vec r-\vec r'|}d\vec r']\psi_{\mu}(\vec r)
+\int d\vec r' V_{xc}(\vec r,\vec r')\psi_{\mu}(\vec r')
=\epsilon_{i}\psi_{\mu}(\vec r)
\end{equation}
The retarded Green's function $G^{R}$ is related to $\psi_{\mu}(\vec r)$ 
through the spectral representation:
\begin{equation}
\label{GR}
G^{R}(\vec r,\vec r';E)
=\sum_{\mu} \frac{\psi_{\mu} (\vec r)\psi_{\mu}(\vec r')^{*}}
{E^{+}-\varepsilon_{i}}
\end{equation} 
where $E^{+}=\lim_{\delta \to 0^{+}} E+i \delta $. 
We expand the wavefunction in terms of a finite set of local 
orbital functions:
\begin{equation}
\label{LCAO5}
\psi_{\mu}(\vec r) \cong \sum_{i}^{N} c_{\mu i}\phi_{i}(\vec r)
\end{equation}
where $\phi_{i}(\vec r)$ are atom-centered and decay rapidly to zero away 
from the corresponding atomic center. We use the symbol $\cong$ to indicate 
the above expansion is exact only for a basis set that is complete. 
\emph{Besides the approximation of single-particle theory, this is the 
only approximation involved in our matrix Green's function 
method}~\cite{Lang82}. Note the choice of the 
basis set can be different in different parts of the system, 
reflecting the different nature of the electronic states in each of them.    
The Schr\"odinger-type equation can be transformed into a generalized 
eigenvalue problem:
\begin{equation}
\sum_{j}H_{ij}c_{\mu j}=\varepsilon_{\mu} \sum_{j} S_{ij}c_{\mu j}
\end{equation}
where $S_{ij}$ is the overlap matrix,
\begin{equation}
S_{ij}=\int d^{3}r \phi_{i}^{*}(\vec r) \phi_{j}(\vec r)
\end{equation}
and
\begin{equation}
\label{Fock}
H_{ij}=\int d^{3}r \phi_{i}^{*}(\vec r) 
[-\frac{1}{2}\nabla^{2}+V_{ext}(\vec r)
 +\int d^{3}\vec r' \frac{\rho(\vec r')}{|\vec r- \vec r'|}]\phi_{j}(\vec r)
 + \int d^{3}r \int d^{3}\vec r' \phi_{i}^{*}(\vec r) 
V_{xc}(\vec r,\vec r')] \phi_{j}(\vec r')
\end{equation}
Instead of solving the above equation,  
we substitute Eq.\ (\ref {LCAO5}) into Eq.\ (\ref{GR}) and write down 
the Green's function in terms of basis function $\phi_{i}$:
\begin{equation}
\label{GRij}
G^{R}(\vec r,\vec r';E)  \cong  
\sum_{\mu} \sum_{i,j}\frac{c_{\mu i} c_{\mu j}^{*}}{E^{+}-\varepsilon_{i}}
\phi_{i}(\vec r) \phi_{j}^{*}(\vec r')  
= \sum_{i,j}G_{ij}^{R}(E) \phi_{i}(\vec r) \phi_{j}^{*}(\vec r')
\end{equation}
where
\begin{equation}
G_{ij}^{R}(E)=\sum_{\mu} \frac{c_{\mu i}c_{\mu j}^{*}}
{E^{+}-\varepsilon_{i}},
\end{equation}
It is straightforward to prove that $G_{ij}^{R}(E)$ satisfies the following 
matrix equation:
\begin{equation}
\sum_{k}(E^{+}S_{ik}-H_{ik})G_{kj}^{R}(E)=\delta_{ij}
\end{equation}
The above equation is an infinite matrix equation involving orbital 
functions centered around every atom of the molecular device. 

Since we are interested only in the Green's function in the ``extended 
molecule'', we 
divide the system into three parts correspond to the left electrode, 
the right electrode and the ``extended molecule'' 
and write down the above equation in block matrix notation:
\begin{equation} 
\left( 
\begin{array}[h]{lll}
E^{+}S_{LL}-H_{LL} & E^{+}S_{LM}-H_{LM} & E^{+}S_{LR}-H_{LR} \\
E^{+}S_{ML}-H_{ML} & E^{+}S_{MM}-H_{MM} & E^{+}S_{MR}-H_{MR} \\
E^{+}S_{RL}-H_{RL} & E^{+}S_{RM}-H_{RM} & E^{+}S_{RR}-H_{RR} \\
\end{array} 
\right) 
\times 
\left( 
\begin{array}[h]{lll}
G_{LL}^{R} & G_{LM}^{R} & G_{LR}^{R} \\
G_{ML}^{R} & G_{MM}^{R} & G_{MR}^{R} \\
G_{RL}^{R} & G_{RM}^{R} & G_{RR}^{R} \\
\end{array} 
\right) 
 = 
\left( 
\begin{array}[h]{lll}
I_{LL} & 0 & 0 \\
0 & I_{MM} & 0 \\
0 & 0 & I_{RR} \\
\end{array} 
\right) 
\label{H3G3}
\end{equation}
The short-range of the local orbital basis means that for any reasonable 
molecule size and electrode spacing, we can neglect the inter-electrode 
block of the Hamiltonian and overlap matrix $H_{LR(RL)}, S_{LR(RL)}$ (this 
also means that we neglect the direct tunneling between the two electrodes) 
and it is straightforward to solve $G_{MM}^{R}$ as:
\begin{eqnarray}
G_{MM}^{R} &=& \{ E^{+}S_{MM}-H_{MM}
-\Sigma_{L}^{R}(E)-\Sigma_{R}^{R}(E)   \}^{-1} \nonumber \\
\Sigma_{L}^{R}(E) &=& (E^{+}S_{ML}-H_{ML}) G^{0;R}_{LL} (E^{+}S_{LM}-H_{LM}), 
\nonumber \\
\Sigma_{R}^{R}(E) &=& (E^{+}S_{MR}-H_{MR}) G^{0;R}_{RR} (E^{+}S_{RM}-H_{RM}), 
\nonumber \\
  G^{0;R}_{LL} &=& (E^{+}S_{LL}-H_{LL})^{-1},  \nonumber \\
G^{0;R}_{RR} &=& (E^{+}S_{RR}-H_{RR})^{-1}.
\label{GMMR}
\end{eqnarray}
Eq.\ (\ref{GMMR}) expresses the Green's function in the molecule 
in terms of the Hamiltonian matrix element in the same region, 
with the coupling to the left and right electrode included rigorously 
in terms of the self-energy operators 
$\Sigma_{L}^{R}(E)$ and $\Sigma_{R}^{R}(E)$. Note again that due to the 
short-range nature of the 
basis set, only the finite block of $G^{0;R}_{LL(RR)}$ is needed for the 
calculation of $\Sigma_{L(R)}^{R}(E)$ corresponding to the orbital basis 
in the left(right) electrode that have non-negligible overlap with the 
orbital basis in the extended molecule. So the calculation 
of $G_{MM}^{R}$ involves matrix operations only on finite matrices.

The matrix self-energy operator can be taken as the matrix elements of a 
non-local operator in real space:
\begin{eqnarray}
\Sigma_{L;ij}^{R}(E) &=&
\int \int d\vec r d\vec r' \phi_{i}^{*}(\vec r)\Sigma_{L}^{R}(\vec r,\vec r')
\phi_{j}(\vec r') \\  
&=& \sum_{mn}(E^{+}S_{ML;im}-H_{ML;im}) G^{0;R}_{LL;mn} 
(E^{+}S_{LM;nj}-H_{LM;nj}) 
\nonumber \\ 
&=& \sum_{mn} 
[\int d\vec r \phi_{i}^{*}(\vec r)[E^{+}-H] \phi_{m}(\vec r)] G^{0;R}_{LL;mn}
[\int d\vec r' \phi_{n}^{*}(\vec r')[E^{+}-H]\phi_{j}(\vec r')]  
\nonumber \\
&=& \int \int d\vec r d\vec r' \phi_{i}^{*}(\vec r) \sum_{mn}[E^{+}-H] 
G^{0;R}_{LL;mn}\phi_{m}(\vec r)\phi_{n}^{*}(\vec r')[E^{+}-H]\phi_{j}(\vec r') 
\nonumber \\
&=&  \int \int d\vec r d\vec r' \phi_{i}^{*}(\vec r) 
\{ [E^{+}-H] G^{0;R}_{LL}(\vec r,\vec r')[E^{+}-H] \} \phi_{j}(\vec r') 
\nonumber   
\end{eqnarray}
with the equation for $\Sigma_{R}^{R}$ taking the same form.

Using the analytic continuation rules of Langreth~\cite{Langreth76}, we 
can write down the corresponding ``lesser'' self-energy matrix as:
\begin{eqnarray}
\Sigma_{L;ij}^{<}(E) 
&=& \int d\vec r d\vec r' \phi_{i}^{*}(\vec r)
\Sigma_{L}^{<}(\vec r,\vec r')\phi_{j}(\vec r') 
\nonumber \\  
&=& \sum_{mn}(E^{+}S_{ML;im}-H_{ML;im})G^{0;<}_{LL;mn}
(E^{+}S_{LM;nj}-H_{LM;nj})  
\end{eqnarray}
Since the left electrode in taken to be in thermodynamic equilibrium, 
we have $G^{0;<}_{LL}(E)=i(G^{0;R}_{LL}(E)-G^{0;A}_{LL}(E))f(E-\mu_{L})$ and 
therefore:
\begin{equation}
\Sigma^{<}_{L;ij}(E) = 
 i(\Sigma^{R}_{L;ij}-\Sigma^{A}_{L;ij})f(E-\mu_{L}) 
 = i\Gamma_{L;ij} f(E-\mu_{L}) 
\end{equation}
where $\Gamma_{L;ij}=\Sigma^{R}_{L;ij}-\Sigma^{A}_{L;ij}$ and 
$f(E-\mu_{L})=\frac{1}{1+e^{(E-\mu_{L})/kT}}$ is the 
Fermi distribution function. The equation for $\Sigma_{R}^{<}$ follows 
by replacing $L$ with $R$ in the above equation. 

Similarly from the kinetic equation Eq.\ (\ref{KKB}), we can write 
down the correlation function in the ``extended molecule'' in terms 
of the basis function $\phi_{i}$ as 
 $G^{<}(\vec r,\vec r';E) 
 = \sum_{ij} G^{<}_{ij}(E) \phi_{i}(\vec r)\phi_{j}^{*}(\vec r') 
 = \int d\vec r'' \int d\vec r''' \sum_{ik} G^{R}_{ik}(E) 
 \phi_{i}(\vec r) \phi_{k}^{*}(\vec r'') \Sigma^{<}(\vec r'',\vec r''';E) 
 \sum_{nj} G^{A}_{nj}(E) \phi_{n}(\vec r''') \phi_{j}^{*}(\vec r')$.  
After rearranging the order of summation and integration and integrating 
over $\vec r''$ and $\vec r'''$, we get 
$G^{<}(\vec r,\vec r';E)  
 = \sum_{ij}\phi_{i}(\vec r) \phi_{j}^{*}(\vec r')
  \sum_{kn} G^{R}_{ik}(E) \Sigma^{<}_{kn}(E) G^{A}_{nj}(E) $. 
Comparing the coefficients of $\phi_{i}(\vec r)\phi_{j}^{*}(\vec r')$, 
we obtain the following matrix equation:
\begin{equation}
\label{GLESS}
G^{<}(E)=G^{R}(E)\Sigma^{<}(E)G^{A}(E)
\label{MKKB}
\end{equation}

If there is no inelastic scattering due to the electron-vibronic coupling, 
we get: 
\begin{equation}
\label{SigL}
\Sigma^{<}(E) =\Sigma^{<}_{L}(E)+\Sigma^{<}_{R}(E)  
= i\Gamma_{L}(E)f(E-\mu_{L})+i\Gamma_{R}(E)f(E-\mu_{R}) 
\end{equation}
since there is no ``lesser'' self-energy operator associated with $V_{xc}$. 
And we can express the correlation function in terms of the distribution 
in each electrodes:
\begin{equation}
G^{<}(E)=i[G^{R}(E)\Gamma_{L}(E)G^{A}(E)]f(E-\mu_{L})
              +i[G^{R}(E)\Gamma_{R}(E)G^{A}(E)]f(E-\mu_{R})
\label{SigLlr}
\end{equation} 
where the products within the brackets are matrix products.  
Every physical observable of interest can be computed from the matrix 
correlation function $G^{<}_{ij}$. In particular, the current density is:
\begin{equation}
\label{IDen}
J(\vec r) = \int dE J(\vec r;E)  = 1/2 \sum_{ij} \int dE G_{ij}^{<}(E)
\lim_{\vec r' \to \vec r}
(\nabla^{'}-\nabla)\phi_{i}(\vec r)\phi_{j}^{*}(\vec r') 
\end{equation}

The terminal current can be calculated numerically by integrating the 
current density $J(\vec r)$ over the boundary surface between the molecule 
and the electrodes or any cross sectional area in the 
``extended molecule'' due to the current continuity in 
the ``extended molecule'' region. But often it is more 
useful to compute the terminal current directly from the matrix Green's 
function and the matrix self-energy operators. This can be achieved by 
defining a current operator, as in Caroli et al.\ ~\cite{Caroli71} and 
Datta ~\cite{Datta95}:
\begin{equation}
I(\vec r,\vec r';E)
 = e/h [H(\vec r)G^{<}(\vec r,\vec r';E)-G^{<}(\vec r',\vec r;E)H(\vec r')]
\end{equation} 
whose diagonal element gives the divergence of the current density:  
\begin{equation}
I(\vec r,\vec r;E)=\nabla \cdot J(\vec r;E) 
\end{equation}

The current over a surface enclosing the ''extended molecule'' is written as:
\begin{eqnarray}
\label{ITerm1}
I_{tot} 
  &=& \int dE \oint_{S}dS \nabla \cdot J(\vec r;E) \nonumber \\
  &=& \int dE \int d\vec r I(\vec r,\vec r;E) \nonumber \\
  &=& e/h \int dE \int d\vec r \sum_{ij} 
  [H(\vec r)G^{<}_{ij}(E) \phi_{i}(\vec r)\phi_{j}^{*}(\vec r) 
  -G^{<}_{ij}(E)\phi_{i}(\vec r)\phi_{j}^{*}(\vec r)H(\vec r)] \nonumber \\
  &=& e/h \int dE \sum_{ij}[H_{ji}G^{<}_{ij}(E)-G^{<}_{ij}H_{ji}] \nonumber \\
  &=& e/h \int dE Tr[HG^{<}(E)-G^{<}(E)H]
\end{eqnarray}

Now again we have transformed the integral over the coordinates to the matrix 
equation involving the Hamiltonian and correlation function matrices. From 
here on the usual derivation using the matrix notation, often used in the 
second-quantized form or with the semi-empirical Hamiltonian matrix, 
can be carried through without change and we get the familiar final form 
for the terminal current in the matrix notation as~\cite{Datta95,Lake97}:
\begin{equation}
\label{ITerm2}
I_{L(R)}=e/h \int dE Tr\{ \Gamma_{L(R)}[f(E-\mu_{L(R)})A(E)+iG^{<}(E)] \}
\end{equation}
where $A(E)=i(G^{R}(E)-G^{A}(E))$ and 
\begin{equation}
I=\frac{e}{h} \int dE 
Tr[\Gamma_{L}(E)G^{R}(E)\Gamma_{R}(E)G^{A}(E)]
[f(E-\mu_{L})-f(E-\mu_{R})]
\end{equation}
where 
\begin{eqnarray}
\Gamma_{L}(E) &=& i(\Sigma_{L}^{R}(E)-[\Sigma_{L}^{R}(E)]^{\dagger}), \\
\Gamma_{R}(E) &=& i(\Sigma_{R}^{R}(E)-[\Sigma_{R}^{R}(E)]^{\dagger})
\end{eqnarray}
Note that in order to represent the coupling to the electrodes as self-energy 
operators, it is essential that the ``extended molecule'' can be described 
by an effective single-particle theory. In other words, the Hamiltonian 
of the ``extended molecule'' in its second-quantized form can be 
diagonalized. If the Hamiltonian describing the the ``extended molecule'' 
contains product over four creation/annihilation operators, the simplified 
description of the electrodes using self-energy operator breaks down and 
more complicated expressions are needed~\cite{MW92,HJBook}.

The above formulae are powerful formal results. Its practical application 
will depend on the existence of efficient algorithms for accurate 
evaluation of the Hamiltonian matrix elements over 
local orbital functions. In practice, a Gaussian-type orbital basis 
is often used~\cite{Wilson87}.
     
\section{Self-Consistent Formulation: Device at Equilibrium}
\label{S4}
\subsection{Model of the molecular device}
The present study considers a single molecule sandwiched between two 
semi-infinite metallic electrodes. In addition, there can be atomic-scale 
features on the metal surface. In the above formulation of our 
matrix Green's function approach, we have not specified the form of the 
Hamiltonian operator. The specific form of the Hamiltonian comes into play 
as we consider the self-consistent formulation. 

The analysis presented here is based on the spin-extension of Kohn-Sham 
density-functional (DF) theory~\cite{Gunn76,Gunn89,Perdew81} 
which requires the solution of effective single-particle, 
Schr\"odinger-like equations with spin-dependent potentials:
\begin{eqnarray}
\label{SKS}
H^{\sigma} \psi_{i}^{\sigma}(\vec r) &=&
[-\frac{1}{2}\nabla^{2}+V_{ion}(\vec r)
 +\int d^{3}\vec r' \frac{\rho(\vec r')}{|\vec r- \vec r'|}
 + V_{xc}^{\sigma}(\vec r)]\psi_{i}^{\sigma}(\vec r)
 =\varepsilon_{i}^{\sigma} \psi_{i}^{\sigma}(\vec r), \nonumber \\
V_{xc}^{\sigma}(\vec r) &=&
\frac{\delta E_{xc}[\rho^{\alpha},\rho^{\beta}]}{\delta \rho^{\sigma}},
\end{eqnarray}
where $\sigma=\alpha,\beta $ and 
\begin{eqnarray}
\rho(\vec r) &=& \rho^{\alpha}+\rho^{\beta}, \nonumber \\
\rho^{\alpha}(\vec r) &=& \sum_{i}n_{i}^{\alpha}
|\psi_{i}^{\alpha}(\vec r)|^{2},  \\
\rho^{\beta}(\vec r) &=& \sum_{i}n_{i}^{\beta}|\psi_{i}^{\beta}(\vec r)|^{2}.
\nonumber
\end{eqnarray}
Here $n_{i}^{\sigma}$ is the occupation number of the spin orbital 
$\psi_{i}^{\sigma}$. For our system, the occupation of the eigenstate is 
governed by the metal Fermi-level $E_{F}$, so at low temperature, 
$n_{i}^{\sigma}=\Theta(\varepsilon_{i}^{\sigma}-E_{F})$. 
In the valence-only calculations, the ionic potential 
$V_{ion}(\vec r)$ is represented by the non-local pseudopotential 
$V_{ps}(\vec r,\vec r')$, and the wavefuctions considered are the 
valence pseudo-orbitals.

The spin-density functional theory (SDF) is the necessary generalization 
in the presence of magnetic field. If the external potential is only of 
electrostatic nature, as in our case due to the atomic nuclei, the 
DF formalism shows that it is possible to determine the system property 
using a functional that depends on the density alone and not the 
spin-densities. The main advantage of the SDF over the DF formalism is that 
the greater flexibility of the SDF formalism introduced by the spin 
dependence allows us to build more of the physics into the approximate 
functional. For example, the SDF formalism can give a reasonable description 
of the bond-breaking in molecules by allowing electrons of different spins 
to have different spatial density distribution. It also provides a better 
description of the open-shell molecules in satisfying the requirement 
(Hund's rule) that a state with a larger spin tends to be favored 
energetically. In addition, spin-orbit coupling  
effects can be included. As a result, the SDF formalism yields 
significantly better results for molecules and solids than the  
spin-unpolarized counterpart~\cite{PYBook,Gunn89}. Our use of the 
SDF formalism here follows that of the unrestricted Hartree-Fock method 
in molecular calculations~\cite{McWeeny89} 
in that we neglect the off-diagonal part of the spin density matrix, 
so the wavefunctions of the $\alpha$- and $\beta$- electrons are decoupled 
from each other except for the spin-dependence in the exchange-correlation 
potential~\cite{Gunn76}(for molecules with even number of electrons 
and singlet spin states, the spin-unrestricted procedure usually 
leads to the same result 
as the spin-restricted one). This breaks the rotational invariance in the 
spin space, and therefore cannot handle situations where the correlation 
between different spin channels are important, e.g., the formation of 
a local magnetic moment at the molecule 
(the Kondo effect)~\cite{Kastner98,Wilk91,MWL93}. These effects 
don't seem to be important, at least when the coupling between the 
molecule and the metal is strong.        

We will consider both the local spin-density approximation (LSDA) 
and its generalized-gradient approximation (GGA), the exchange-correlation 
energy of which takes the following form:
\begin{eqnarray}
\label{GGA}
E_{xc}[n_{\alpha},n_{\beta}]
& = &\int d^{3}r f(\rho_{\alpha},\rho_{\beta},
\gamma_{\alpha \alpha},\gamma_{\alpha \beta},\gamma_{\beta \beta}), 
\nonumber \\
& & \gamma_{\alpha \alpha}=|\nabla n_{\alpha}|^{2},\\
& & \gamma_{\alpha \beta}=\nabla n_{\alpha} \cdot \nabla n_{\beta},
\nonumber \\
& & \gamma_{\beta \beta}=|\nabla n_{\beta}|^{2} \nonumber 
\end{eqnarray}

\subsection{The self-consistent formulation}
 
The starting point of computing the nonlinear transport characteristics 
of the molecular device, is then to compute the self-consistent charge 
and potential distribution of the device at equilibrium. 
This problem is equivalent 
to calculating the electronic structure of the 
metal-molecule-metal junction, which is the generalization of the 
familiar chemisorption problem where we consider the adsorption of a 
single isolated molecule onto one metal surface. Chemisorption 
is an important part of surface physics/chemistry and substantial 
theoretical studies exist in the literature, using the 
wavefunction~\cite{Lang78} or Green's function 
formulation~\cite{Scheffler94,Feibel87,Feibel92,Pollmann88,Ingle96,Scheffler91,Scheffler92}, 
and working in real space~\cite{Lang78,Ingle96} or in finite basis 
expansion~\cite{Feibel87,Feibel92,Pollmann88} or combinations of the 
two~\cite{Scheffler94,Scheffler91,Scheffler92}. Our method is similar 
to these methods in many details. However, the uniqueness of our approach 
is that we use the finite basis expansion from the beginning, which allows 
us to separate the entire system naturally into different parts 
according to their geometrical arrangement, the treatment of which 
can then be improved independent of each other. This allows us to 
use existing, well-established techniques for treating the 
component systems---the molecules and the surface or the bulk---
directly in the study of molecular devices with little change.   

As explained in the above, instead of solving the Kohn-Sham 
equation directly, we approximate the single-particle 
wavefunction by an expansion in a finite set of local orbital basis 
functions and solve the matrix Green's function $G_{ij}^{\sigma}(Z)$ as 
the solution of the matrix equation:
\begin{equation}
\label{MGF}
\sum_{k}(ZS_{ik}-H^{\sigma}_{ik})G_{kj}^{\sigma}(Z)=\delta_{ij}
\end{equation}
Here $Z$ is an arbitrary complex energy. From the spectral representation 
relation: 
\begin{equation}
\label{Spect}
G_{ij}^{\sigma}(Z)=\sum_{\mu} \frac{c_{\mu i}^{\sigma}c_{\mu j}^{\sigma *}}
{Z-\varepsilon_{i}},
\end{equation}
we can obtain the density matrix $\rho_{ij}^{\sigma}$, 
which is defined as 
\begin{equation}
\label{dmatrix}
\rho_{ij}^{\sigma}=\sum_{\mu}n^{\sigma}_{\mu}
c_{\mu i}^{\sigma}c_{\mu j}^{\sigma *}
\end{equation}
by performing the following contour integration in the 
complex energy plane:
\begin{equation}
\label{contour}
\rho_{ij}^{\sigma}=\frac{1}{2\pi i}\int_{C}dZ G_{ij}^{\sigma}(Z),
\end{equation}
where the integration contour encloses the real energy axis from the 
energy of the lowest occupied state up to the Fermi 
energy $E_{F}$~\cite{Lang82,Dede82}. 
A typical integration contour  is shown in Fig.\ (\ref{Fig32}). The 
advantage of integrating along the complex energy contour is that 
by moving away from the real energy axis, the sharp features in the 
density-of-states are smoothed, therefore allowing for accurate integration 
with a small integration mesh~\cite{Dede82}. 

The electronic charge density $\rho^{\sigma}(\vec r)$, which is needed 
for the evaluation of $H_{ij}^{\sigma}$, can be computed according to
\begin{equation}
\label{CDens}
\rho^{\sigma}(\vec r)=\sum_{i,j} 
\rho_{ij}^{\sigma} \phi_{i}(\vec r) \phi_{j}^{*}(\vec r)   
\end{equation}
We can also obtain the local density of states (DOS), defined as 
\begin{eqnarray}
\label{LDOS}
n^{\sigma}(\vec r,E) & = & \sum_{\mu} |\psi^{\sigma}_{\mu}(\vec r)|^{2}
\delta(E-\varepsilon_{\mu}) \nonumber \\
& = & \sum_{\mu} \sum_{i,j} c_{\mu i}^{\sigma}c_{\mu j}^{\sigma *}
\phi_{i}(\vec r) \phi_{j}^{*}(\vec r)\delta(E-\varepsilon_{\mu}),
\end{eqnarray}  
from the following equation:
\begin{equation}
\label{G2DOS}
n^{\sigma}(\vec r,E)=-\frac{1}{\pi} \lim_{\delta \to 0^{+}}\sum_{i,j}
Imag[G_{ij}^{\sigma}(E+i\delta)] \phi_{i}(\vec r) \phi_{j}^{*}(\vec r)
\end{equation}
and the total density of states 
\begin{equation}
\label{TDOS}
n^{\sigma}(E)=\int d^{3}r n^{\sigma}(\vec r,E)=-\frac{1}{\pi} 
\lim_{\delta \to 0^{+}}\sum_{i,j}Imag[G_{i,j}^{\sigma}(E+i\delta)]S_{ji}
=-\frac{1}{\pi} Tr\{ Imag[GS] \}
\end{equation}
We can also obtain the projection of the total density of states into 
the molecular region as
\begin{equation}
\label{PDOS}
n^{\sigma}_{Mol}(E)=-\frac{1}{\pi} Tr\{ Imag[GS]|_{Mol} \}
\end{equation}
where the trace is taken with respect to the orbital indices of the 
molecule. The transmission 
coefficient through the molecule is determined from:
\begin{eqnarray}
\label{TE}
T(E) &=& Tr\{ \Gamma_{L}G^{R} \Gamma_{R}G^{A} \}, \nonumber \\
\Gamma_{L} &=& i(\Sigma^{R}_{L}-(\Sigma^{R}_{L})^{\dagger}), \\
\Gamma_{R} &=& i(\Sigma^{R}_{R}-(\Sigma^{R}_{R})^{\dagger}). \nonumber 
\end{eqnarray}
From here on, we avoid writing down explicitly the spin indices unless 
otherwise noted, with the understanding that we need to solve the Green's 
function corresponding to each spin direction and the total transmission is 
the summation over the two spin channels.
      
The self-consistent computation is greatly simplified by realizing that 
due to the metallic screening within the electrode, the charge and 
potential perturbation induced by the adsorption of the molecule extends 
over only a finite region into the electrodes, practically only the region 
including the surface metal atoms closest to the molecule~\cite{BT99}. 
The charge and potential distributions beyond this region are the same 
as that of the bimetallic interfaces without the molecules. It is this 
region---the molecule plus the 
perturbed surface atoms--- that enters the above matrix Green's  
formulation and is called the ``extended molecule''. 
The computation of the Green's function of the left and 
right electrodes then depends only on the charge and potential 
distribution of the bimetallic interfaces, which can be calculated separately 
and need to be calculated only once.  
Note that in this formulation we have neglected the long-range charge 
perturbation to the metal surface due to the presence of nonzero electric 
field in the ``extended molecule''. This long-range charge perturbation 
could be important if we want to compute quantities that increase with 
distance such as the dipole moment associated with the adsorbed molecule. 
Since we are mostly interested in the electrostatic potential in the 
``extended molecule'' region  which is inversely proportional to distance, 
no significant error will be introduced by neglecting these long-range 
charge perturbations~\cite{Lang78,BT99}.

The device characteristic is determined only by the electronic processes 
in the finite ``extended molecule'' region. In the MGF formalism, the 
effects of the contacts enter in two different 
ways: (1) as the infinite electron source and drain, the contacts inject 
electrons into and absorb electrons from the molecule, the occupation of 
the single-particle states within which is set by the electrochemical 
potential of the contacts. The self-energy operators describe this 
interaction between the molecular states with the continuum of states in 
the contacts, which of course depends on the band structure of the metal. 
(2) as the boundary condition, the charge and potential distribution in the 
``extended molecule'' must join the charge and potential 
distribution deep within the interior of the contacts. Since the 
metallic electrodes are described well by the 
local-density-approximation, we assume only the exchange-correlation term 
in the ``extended molecule'' region may have a nonlocal form. So only the 
determination of the electrostatic potential is constrained by the 
long-range coulomb interaction such that it joins to the bulk value at 
regions beyond the ``extended molecule''. The exchange-correlation 
potential depends only on the charge distribution in the 
``extended molecule'' region, so it is irrelevant when we discuss the 
constraint of the boundary condition on potential imposed by the presence 
of the electrodes. 

Note that the unperturbed part of the contacts are the semi-infinite 
crystal with the perturbed surface atoms \emph{removed}. Its Green's 
function $G_{LL(RR)}^{0;R}$ can be calculated from that of the semi-infinite 
crystal using the ``reduced space'' idea of Williams, Feibelman and 
Lang~\cite{Lang82}, which is essentially a two-component version of our 
MGF equation (Eq.\ (\ref{H3G3})). In this 
approach, we deal with the following two-by-two block matrix equation:
\begin{equation} 
\label{H2G2}
\left( \begin{array}{cc}
E^{+}S_{LL(RR)}-H_{LL(RR)} & E^{+}S_{L(R)P}-H_{L(R)P} \\
E^{+}S_{PL(R)}-H_{PL(R)} & E^{+}S_{PP}-H_{PP}  \\
\end{array} \right) \times 
\left( \begin{array}{cc}
G_{S;LL(RR)}^{R} & G_{S;L(R)P}^{R} \\
G_{S;PL(R)}^{R} & G_{S;PP}^{R} \\ 
\end{array} \right)
=
\left( \begin{array}{cc} 
 I_{S;LL(RR)} & 0 \\
 0 & I_{S;PP} \\
\end{array} \right)
\end{equation}
where we have separated the semi-infinite surface into two parts with 
$P$ denoting the region of the \emph{semi-infinite} surface that is 
included in the ``extended molecule'' and $L(R)$ denoting the region of 
the \emph{semi-infinite} surface corresponding to the unperturbed 
part of the left (right) electrode. From the above equation, we get:
\begin{eqnarray}
\label{GLR}
G^{0;R}_{LL(RR)} & = & (E^{+}S_{LL(RR)}-H_{LL(RR)})^{-1} \\
   & = & G_{S;LL(RR)}^{R}- G_{S;L(R)P}^{R}
\{ G_{S;PP}^{R} \}^{-1}G_{S;PL(R)}^{R} \nonumber 
\end{eqnarray}
where again only matrix operations over finite matrices are involved due 
to the shore-range nature of the basis functions. 

The problem associated with the long-range electrostatic potential is 
commonly attacked using the scattering theory of electronic structure 
by writing down the Hamiltonian of the total system as a summation of 
an unperturbed part and a localized 
perturbation~\cite{Lang78,Scheffler94,Pollmann88,Scheffler92}, 
which is then solved using the Dyson equation, 
\begin{eqnarray}
H &=& H^{0}+\Delta V^{eff}
=-\frac{1}{2}\nabla^{2}+V^{0}(\vec r)+\Delta V^{eff}, \\
G &=& G^{0}+G^{0}\Delta V^{eff}G
\label{SCP}
\end{eqnarray}
where the change in the effective potential is:
\begin{equation}
\label{DV}
\Delta V^{eff}[\rho] = \Delta V^{ion}
+\int d^{3}\vec r^{'} \frac{[\rho(\vec r')-\rho^{0}(\vec r')]}
{|\vec r-\vec r'|}+V_{xc}[\rho(\vec r)]
-V_{xc}[\rho^{0}(\vec r)] 
\end{equation}
In the above formulation, the ``reference potential'' $V^{0}$ and the 
``reference charge density'' $\rho^{0}$ are those obtained from the 
self-consistent calculation of the bimetallic interfaces 
(without the molecule). In this way, $\Delta V^{eff}$ is nonzero only 
within the ``extended molecule''. $\Delta V^{ion}$ is the ionic potential 
due to the atomic ions in the isolated molecule plus any changes 
in the ionic potential that may be caused by the surface reconstruction 
and/or the modification of the molecular structure upon the adsorption onto 
the surfaces (any such structural relaxation is taken as occuring only 
in the ``extended molecule''). 
  
Since the MGF formalism allows us to calculate the Green's function of the 
``extended molecule'' given the potential within the same region, we want 
to transform the above equation into a form that better suits our purpose.
Since the exchange-correlation potential in the ``extended molecule'' 
doesn't depend on the charge distribution outside this region, we don't 
need to include it in the choice of $V^{0}$. This will also allow us to 
use different approximation schemes for the exchange-correlation 
potential in different part of the total system. Instead of calculating 
the change in the potential, we take a more constructive approach to 
get the potential directly and write down the Hamiltonian of the 
``extended molecule'' as:
\begin{eqnarray}
\label{Veff}
H^{\sigma} &=& -\frac{1}{2}\nabla^{2}+V^{0}(\vec r)+V^{ion}(\vec r)-V^{ion,0}+
\int d^{3}\vec r^{'} \frac{[\rho(\vec r')-\rho^{0}(\vec r')]}
{|\vec r-\vec r'|}+V_{xc}^{\sigma}(\vec r), \nonumber \\ 
  &=&  -\frac{1}{2}\nabla^{2}+V^{ext,0}(\vec r)+V^{ion}(\vec r)
+\int d^{3}\vec r^{'} \frac{\rho(\vec r')}{|\vec r-\vec r'|}
+V_{xc}^{\sigma}(\vec r), 
\end{eqnarray}
where 
\begin{equation}
\label{Vext}
V^{ext,0}(\vec r)=V^{0}(\vec r)-V^{ion,0}
-\int d^{3}\vec r^{'} \frac{\rho^{0}(\vec r^{'})}{|\vec r-\vec r^{'}|}
\end{equation}
Here $V^{ext,0}(\vec r)$ represents the electrostatic potential due to the 
charge (ionic and electronic) distribution in the contact region, .i.e., 
outside the ``extended molecule''. Since the contact region is 
charge neutral as a whole, $V^{ext,0}(\vec r)$ decays fast away from the 
contact region. If sufficient number of metal surface atoms are included 
into the ``extended molecule'', $V^{ext,0}(\vec r)$ will be negligible 
within the molecule part.   

In principle, a separate self-consistent calculation of the bimetallic 
interface without the adsorption of the molecule needs to be 
performed~\cite{FS85}. For electrodes 
made from the same material, the results are practically the superposition 
of two separate surface calculations. For electrodes made from different 
materials, boundary conditions on the electrostatic potential will be 
imposed which line up the Fermi-levels of the two electrodes~\cite{FS85}. 
Such a calculation will give us both the self-consistent charge and potential 
distribution in the region that we take to be beyond the range of 
admolecule perturbation, from which we can also get the corresponding 
surface Green's function $G_{LL}^{0}$ and $G_{RR}^{0}$. However, their 
exact value will depend on the particular model of the 
surface \emph{and} the choice of the boundary 
between the perturbed and unperturbed surface region. Since the exact 
nature of the contact is almost never known in most experiments on 
molecular devices, and since the change in the electronic structure of 
the molecule is mainly determined by the local interaction between the 
molecule and the neighbor surface atoms that are included in the 
``extended molecule'', a good description of the contact can be obtained 
if we calculate the potential and the Green's function of the unperturbed 
part of the electrodes using values obtained from bulk calculations 
(for electrodes made from different materials, an additional linear 
potential term corresponding to the work function difference across 
the electrodes will need to be added to the bulk values for electrostatic 
potential calculations).   
This greatly simplifies the calculation while allowing meaningful 
quantitative comparison between theory and experiments. In particular, 
we note that satisfactory description of the charge and potential 
distribution in the bulk can be obtained without performing fully 
self-consistent calculations, for example, using the 
overlapping-atomic-potential model~\cite{Louie84}, muffin-type 
approximations~\cite{Andersen75} or tight-binding parameters. This 
approximate treatment of contacts can always be improved without 
affecting other parts of the calculation.    

Given the Hamiltonian of Eq.\ (\ref{Veff}), the matrix elements entering 
the Fock matrix in Eq.\ (\ref{Fock}) can then be separated into the 
core, Coulomb and exchange-correlation matrix as usual in the molecular 
density-functional calculations~\cite{Pople92}:
\begin{equation}
\label{FockM}
H^{\sigma}_{ij}=H_{ij}^{core}+J_{ij}+F_{xc;ij}^{\sigma}
\end{equation}
where $H_{ij}^{core}$ is the summation of the kinetic-energy, the ionic 
potential and the ``external potential'' matrices, and $J_{ij}$ is the 
Coulomb matrix:
\begin{eqnarray}
\label{Coul}
J_{ij}=\sum_{kl} \rho_{kl} (ij|kl), \nonumber \\
\rho_{kl}=\rho_{kl}^{\alpha}+\rho_{kl}^{\beta}
\end{eqnarray}
where we have used the conventional notation for the electron repulsion 
integrals (ERI). For exchange-correlation energy of the functional form as 
Eq.\ (\ref{GGA}), the exchange-correlation potential and their matrix 
elements are given by:
\begin{eqnarray}
\label{Vxc}
V_{xc}^{\alpha}[\rho^{\alpha},\rho^{\beta}] &= &
\frac{\partial f[\rho^{\alpha},\rho^{\beta}]}{\partial \rho^{\alpha}}, \\
F_{xc;ij}^{\alpha} &= & \int 
V_{xc}^{\alpha} \phi_{i}^{*}(\vec r)\phi_{j}(\vec r) d^{3}\vec r
\end{eqnarray}
and similarly for $V_{xc}^{\beta}$ and $F_{xc;ij}^{\beta}$.

At this point the outline of the calculation of the equilibrium property 
of molecular devices using the self-consistent matrix Green's function 
method is complete: \\
(1) A judgment is made as to the set of atomic sites on the electrode 
surfaces which are included into the ``extended molecule''. \\
(2) A judgment is made as to the set of atomic sites on the electrode 
surfaces which are considered to be coupled to the ``extended molecule''. \\
(3) A basis set ${\phi_{i}(\vec r)}$ is chosen for the atoms within the 
``extended molecule''. \\
(4) The Hamiltonian matrix of the unperturbed part of the electrodes and also 
the coupling between them and the ``extended molecule'' is determined from 
either the self-consistent bulk charge/potential distribution or from an 
approximate construction such as the tight-binding model. 
The corresponding surface Green's function 
and self-energy matrix is then calculated at a preselected numerical 
energy integration mesh. \\
(5) The ``external potential'' $V^{ext,0}$ and its matrix elements 
$V^{ext,0}_{ij}$ are calculated either from a self-consistent surface/bulk 
calculation or by approximate construction. Also calculated are the 
matrix elements of 
the kinetic energy operator and the ionic potentials. Together they give the 
core Hamiltonian matrix. \\
(6) An initial guess is made for the spin density matrix 
$\rho_{MM}^{\sigma}$ of the ``extended molecule''. A natural choice 
for the initial guess is the density matrix of the 
free ``extended molecule''. An even  better one is the density matrix of 
the free ``extended molecule'' under the action of 
the ``external potential'' $V^{ext,0}$. \\ 
(7) Calculate the Coulomb and exchange-correlation part of the Fock matrix. 
For Gaussian-type orbital (GTO) basis, the Coulomb matrix elements can be 
calculated analytically. The existence of efficient algorithms for this 
operation is the main strength of GTO over other choices of local atomic 
orbital functions such as Slater-type orbitals (STO). The calculation 
of the exchange-correlation matrix elements must be performed 
numerically over the $3$-d molecular volume due to the rather 
complicated form of the exchange-correlation 
potential (Eq.\ (\ref{GGA})). \\  
(8) Eq.\ (\ref{MGF}) is then solved to obtain $G_{MM}^{\sigma}(Z)$ for 
each $Z$ of the numerical integration mesh. The corresponding density-matrix 
$\rho_{MM}^{\sigma}$ is then 
recalculated via the numerical integration over the complex energy 
contour using Eq.\ (\ref{contour}). \\
(9) Repeat step ($7$) and ($8$) until the input density or Fock matrix 
agrees 
with the output density or Fock matrix within a preset range. Due to the 
long-range nature of Coulomb interaction and the ``heterogeneous'' character 
of the ``extended molecule'', strong oscillations often occur in the 
iteration process, therefore acceleration methods for self-consistent 
convergence are generally needed for the self-consistent process 
to converge~\cite{DZ83}. \\
(10) The electronic density of states, charge transfer, electrostatic 
potential and transmission coefficients are then calculated and analyzed.

Note the above procedures are almost identical to that of calculating the 
electronic structure of the free ``extended molecule'' under some external 
potential $V^{ext,0}$, the only exception being that at each iteration, 
instead of diagonalizing the Fock matrices, we integrate along the complex 
energy contour to obtain the density matrix $\rho_{ij}$ for the next 
iteration (see the program flow-chart of Fig.\ \ref{Fig33}). 
In particular, given the density matrix $\rho_{ij}$, the computation 
of the Hamiltonian matrix $H_{ij}^{\sigma}$ is the 
same as that of the free ``extended molecule'' under the ``external'' 
potential $V^{ext,0}(\vec r)$ which, like the self-energy operator, depends 
only on the charge and potential distributions obtained from 
the separate surface/bulk calculations and needs to be calculated only 
once. Since the above step is the computationally most demanding one in our 
self-consistent calculation, this allows us to take advantage of the full 
repertoire of molecular electronic structure calculations. In practice, 
this step can be replaced by call to any existing high-quality 
quantum-chemical software such as Gaussian 98~\cite{G98}.            

\section{Self-Consistent Formulation: Device out of equilibrium}  
\label{S5}
\subsection{The meaning of the voltage drop}
Given the charge $\rho^{0}$ and the electrostatic potential $V^{0}_{es}$ 
distribution of the device at equilibrium, applying an external bias 
will in general change the charge density and correspondingly the 
electrostatic potential (see Fig.\ \ref{Fig34}), which are related 
through the Poisson equation:
\begin{eqnarray}
\nabla^{2} V_{es}(\vec r)=\rho(\vec r), \\
\nabla^{2} V_{es}^{0}(\vec r)=\rho^{0}(\vec r),
\end{eqnarray}
or 
\begin{equation}
\nabla^{2} \delta V_{es}(\vec r)=\delta \rho(\vec r)
\end{equation}
where $\delta V_{es}=V_{es}-V_{es}^{0}$ and 
$\delta \rho=\rho-\rho^{0}$. 
The boundary condition on $\delta V_{es}$ 
is then:
\begin{equation}
\label{DelV}
\delta V_{es}(\vec r) \vert_{L}=V_{L}, \ \delta V_{es}(\vec r) \vert_{R}=V_{R} 
\end{equation} 
where $V_{L(R)}$ is the change of the electrostatic potential in the 
interior of the left (right) electrodes when current is flowing. 
$V_{R}-V_{L}=W$ could in general be different from $V$ while 
$\mu_{R}-\mu_{L}=eV$~\cite{Lang92}.  
$\delta V_{es}$ describes the ``voltage drop'' across the molecule. 
The solution of Eq.\ (\ref{DelV}) can be 
separated into two parts, which also makes its physical meaning clear:
\begin{equation}
\label{DelVes}
\delta V_{es}(\vec r)= V_{bias}(\vec r)
+\int d\vec r' \frac{\delta \rho(\vec r')}{|\vec r-\vec r|}
\end{equation}
where $V_{bias}(\vec r)$ is the solution of 
\begin{eqnarray}
\nabla^{2} V_{bias}(\vec r)=0, \nonumber \\
V_{bias}(\vec r) \vert_{L}=V_{L},  V_{bias}(\vec r) \vert_{R}=V_{R} 
\end{eqnarray}
Note this is of the same form as we will obtain for the bare biased 
bimetallic interfaces. As is the case of the bare bimetallic interface, 
$V_{bias}$ is linear. The final electrostatic potential is 
then given by:
\begin{equation}
\label{Vest}
V_{es}(\vec r)=V_{es}^{0}(\vec r)+V_{bias}(\vec r)
+\int d\vec r' \frac{\delta \rho(\vec r')}{|\vec r-\vec r|}
\end{equation}
where $\delta \rho$ is the charge \emph{redistribution} in the ``extended 
molecule'' due to the applied bias. Written in this form, we see that 
$V_{bias}$ doesn't depend on the property of the ``extended molecule'', but 
only reflects the change in the boundary condition due to the applied bias 
and that is why we have used the subscript $V_{bias}$. 
The charge transfer $\delta \rho$ is the charge redistribution within 
the ``extended molecule'' induced by this ``applied bias'' $V_{bias}$.   

In this way, we can reach some qualitative conclusions about the voltage 
drop and the current-voltage characteristics of the molecular device. 
Obviously if the geometry of the molecule and the arrangement of the 
electrodes are such that the system is perfectly symmetric with respect to 
mirror reflection across the plane crossing the middle of the molecule, 
the current-voltage characteristic will be symmetric with respect to the 
bias. This can be seen from the above formula where changing the 
sign of $V$: 
$V \to -V$ is equivalent to changing the coordinate $z \to - z$ where we 
assume current flows along $z$ axis. 
Since the choice of left or right is simply a matter of convention, the 
magnitude of the current flowing through the device can't change. Again in 
the symmetric case, since the charge neutrality is maintained in the 
``extended molecule'' $\int d\vec r \delta \rho(\vec r)=0$, 
$\delta \rho(\vec r)$ must be zero at the plane crossing the middle, so 
we expect the voltage drop will be close to the linear form in the middle 
of the molecule, and deviate from it as we move away from the middle 
towards the boundary set by the metal surface. 

\subsection{The self-consistent formulation}

Note the equilibrium electrostatic potential $V_{es}^{0}$ is related to 
the equilibrium charge distribution $\rho^{0}$ through the following 
equation: 
\begin{equation}
V_{es}^{0}(\vec r)=V^{ext,0}(\vec r)+V^{ion}(\vec r)+ 
\int  d\vec r' \frac{ \rho^{0}(\vec r')}{|\vec r-\vec r|}
\label{Ves0}
\end{equation}
From Eq.\ (\ref{Vest}), the electrostatic potential at nonzero 
bias $V_{es}^{0}$ is thus: 
\begin{equation}
V_{es}(\vec r)=V_{bias}(\vec r)+V^{ext,0}(\vec r)+V^{ion}(\vec r)+ 
\int  d\vec r' \frac{ \rho(\vec r')}{|\vec r-\vec r|}
\end{equation}
Defining $V^{ext}(\vec r)=V_{bias}(\vec r)+V^{ext,0}(\vec r)$, we can write 
down the Hamiltonian of the molecular device at nonzero bias in the same 
form as that at equilibrium (Eq.\ (\ref{Veff})): 
\begin{equation}
\label{HV}
H^{\sigma}=  -\frac{1}{2}\nabla^{2}+V^{ext}(\vec r)+V^{ion}(\vec r)
+\int d^{3}\vec r^{'} \frac{\rho(\vec r')}{|\vec r-\vec r'|}
+V_{xc}^{\sigma}(\vec r), 
\end{equation}
Again $V^{ext}$ can be regarded as the electrostatic potential due to the 
charge distribution outside the ``extended molecule''. The addition of 
$V_{bias}$ reflects the change in the boundary condition in this 
electrostatic potential~\cite{Datta98,Ratner00,EK00}.  

For device out of equilibrium, the central quantity is the matrix 
correlation function $G^{<}_{ij}(E)$, which is computed using the 
KKB equation: 
\begin{equation}
G^{<}(E)=i[G^{R}(E)\Gamma_{L}(E)G^{A}(E)]f(E-\mu_{L})
              +i[G^{R}(E)\Gamma_{R}(E)G^{A}(E)]f(E-\mu_{R})
\label{NEGL}
\end{equation}  
The self-consistent calculation then 
proceeds by computing the input density matrix to the next iteration 
from the correlation function computed in the current iteration:
\begin{eqnarray}
\rho(\vec r) &=& 
\sum_{ij}\rho_{ij}\phi_{i}(\vec r)\phi_{j}^{*}(\vec r) \nonumber \\ 
        &=& \int \frac{dE}{2\pi i}G^{<}(\vec r,\vec r;E) \nonumber \\ 
        &=& \sum_{ij} \int \frac{dE}{2\pi i}G^{<}_{ij}(E) 
            \phi_{i}(\vec r)\phi_{j}^{*}(\vec r)
\end{eqnarray}
or in the matrix formulation:
\begin{equation}
\label{DMNE}
\rho_{ij}=\int \frac{dE}{2\pi i}G^{<}_{ij}(E)
\end{equation}
\emph{The density matrix is nothing but the energy integration of the 
matrix correlation function}. To see better the physical meaning of this, 
we divide both sides of Eq.\ (\ref{NEGL}) by:
\begin{equation}
\label{NEA}
A(E)= i(G^{R}(E)-G^{A}(E)) 
    = G^{R}(E)(\Gamma_{L}(E)+\Gamma_{R}(E))G^{A}(E)   
\end{equation}
and get:
\begin{eqnarray}
\frac{-iG^{<}(E)}{A(E)} &=& 
      f(E-\mu_{L})\gamma_{L}+f(E-\mu_{R})\gamma_{R}, \nonumber \\
\gamma_{L} &=& G^{R}(E)\Gamma_{L}(E)G^{A}(E)/A, \\
\gamma_{R} &=& G^{R}(E)\Gamma_{R}(E)G^{A}(E)/A. \nonumber
\end{eqnarray}
where the division of matrices is defined such that 
$[\frac{A}{B}]_{ij}=\frac{A_{ij}}{B_{ij}}$. 
Since the correlation function $-iG^{<}$ describes the number of 
electrons at energy $E$ and the spectral function $A$ describes the density 
of states at energy $E$, the above equation essentially says that the 
probability of the states at energy $E$ being occupied in the molecule 
equals the probability of it being occupied in the left electrode 
multiplying the escape rate $\gamma_{L}$ from the left electrode to 
the molecule plus the probability of it being occupied in the right 
electrode multiplying the escape rate $\gamma_{R}$ from the right 
electrode to the molecule.
  
Note the integration contour appearing in Eq.\ (\ref{DMNE}) is along the 
real energy axis rather than in the complex energy plane. Unlike the 
retarded Green's function, the correlation Green's function $G^{<}(E)$ is 
not analytic away from the real energy axis. This would have increased 
significantly the computational cost of the energy integration. However, 
for energy sufficiently lower than both $\mu_{L}$ and $\mu_{R}$, say 
$E < \mu_{L(R)}-10kT$, we have $f(E-\mu_{L(R)}) \approx 1$ and 
Eq.\ (\ref{NEGL}) reduces to: 
\begin{equation}
G^{<}(E)=i[G^{R}(E)\Gamma_{L}(E)G^{A}(E)]
        +i[G^{R}(E)\Gamma_{R}(E)G^{A}(E)]
\end{equation}     
Comparing with Eq.\ (\ref{NEA}), we have: 
\begin{equation}
G^{<}(E)=iA(E)=-2iImag[G^{R}(E)]
\end{equation}
Consequently the integration over the energy in Eq.\ (\ref{DMNE}) can be 
split into two parts: 
\begin{equation}
\label{DMNE2} 
\rho_{ij}=\frac{1}{2\pi i} \int_{C}dZ G_{ij}(Z) 
         +\frac{1}{2\pi i} \int_{E_{min}}^{E_{max}} dE G^{<}_{ij}(E)
\end{equation}
where the first term represents integration along the same complex contour as 
that in Fig.\ (\ref{Fig32}) with the upper energy cutoff replaced 
by $E_{min}$ and the second term represents integration along real 
energy axis from $E_{min}$ to $E_{max}$. Here $E_{min(max)}$ is chosen 
such that for $E<E_{min}$, $f(E-\mu_{L(R)}) \approx 1$ and 
for $E>E_{max}$,$f(E-\mu_{L(R)}) \approx 0$. The integration along the 
real energy axis can be performed using fine integration grids since the 
range of integration is not much larger than $eV$.  

It is clear from the above discussion that the only difference 
between the self-consistent formulation for device out of equilibrium 
and for device at equilibrium lies in the way of calculating the density 
matrix (Eq.\ (\ref{DMNE2})). At nonzero bias, the retarded Green's function 
alone is no longer adequate for describing the observable property of 
the system. Instead, at each iteration, we need to calculate the 
correlation function from the retarded Green's function, the energy 
integration of which gives the input density matrix for the next iteration. 
\emph{All other procedures remain the same}. After self-consistency is 
achieved, we can calculate the terminal current using Eq.\ (\ref{ITerm1}) or 
Eq.\ (\ref{ITerm2}) and also the current density using Eq.\ (\ref{IDen}). 
As a result, calculating current transport in molecular electronic devices is 
no more difficult than solving a molecular chemisorption problem, 
both conceptually and computationally.  

\section{Conclusion}

Modeling transport at the molecular scale requires a microscopic knowledge 
of the electronic structure of both the molecule and the surface which has 
to be considered in the context of an open system exchanging particles and 
energy with the external contacts. This complicates substantially the 
computational procedure and raises new questions about the theoretical basis. 
Although a first-principles theory of transport at the molecular scale 
doesn't yet exist, we have shown that the Non-equilibrium Green's Function 
Formalism of quantum transport combined with the density-functional theory 
of electronic structure provides a sound basis on which further works may  
be based. 

Since the electronic process in such molecular devices is mainly 
determined by the interaction between the molecule and the surface atoms 
closest to it, it is highly desirable to separate the treatment of the 
physical processes in this ``extended molecule'' region from that in 
the rest of the system since in general the exact atomic geometry of 
the contact is not known. In addition, as we have described, the 
treatment of the contact is best chosen to make the 
description of the physical processes in the ``extended molecule'' 
consistent. This is conveniently dealt with within the matrix Green's 
function approach using expansion in finite set of local orbital functions, 
where the effect of the contacts is taken into account using self-energy 
operators $\Sigma_{L(R)}$. 
In addition, by introducing an ``external potential'' $V^{ext}$ which 
describes the electrostatic potential due to the charge distribution 
outside the ``extended molecule'', we have shown rigorously that the 
electronic and transport property of the molecular device can be 
determined from the electronic processes in the ``extended molecule'' 
alone, given the knowledge of $V^{ext}$ and $\Sigma_{L(R)}$. This 
allows the highly 
accurate and efficient techniques developed for molecular electronic 
structure computation to be used for studying transport through molecules 
with little change. We believe such an approach will greatly accelerate 
theoretical/computational research on molecular electronic devices. 

Besides the computational advantage, the use of the local orbital basis 
sets and the techniques of molecular electronic structure theory will also 
greatly facilitate the interpretation of the result of computation using 
the language of qualitative molecular orbital theory which has provided 
the rationalization of the intuitive picture of bonding and orbital 
interactions in molecules~\cite{Hoffmann88,ABW85}. Traditionally such 
discussion has relied on semi-empirical theories such as the 
Extended H\"{u}ckel type of theory. The advancement of density-functional 
theory in quantum chemistry has made the use of self-consistent 
Kohn-Sham orbitals in such molecular orbital theory highly 
desirable~\cite{Hoffmann99,Baerends}. For molecular electronic 
devices, the orbitals involved are those of the molecule and the surface 
atoms closest to it. Much of the physics can then be understood in terms 
of the orbital 
interactions \emph{after the effect of self-consistent charge transfer and 
potential distribution has been included}. Such an approach has been taken 
recently by the present authors to study the equilibrium property of the 
molecular device formed by the phenyldithiolate molecule bridging two 
gold contacts~\cite{Xue011}. Further work on nonlinear transport 
through molecules is under investigation~\cite{Xue012}.            

{\bf Acknowledgements}: This paper is based on a Ph.\ D.\ Thesis by Y.\ Xue 
at Purdue University. The work at Purdue is supported by the National 
Science Foundation under Grant No.\ 9708107-DMR. The work at Northwestern 
is supported by The DARPA Molectronics program, The DoD-DURINT program 
and The NSF Nanotechnology Initiative. 

\emph{Notes added after proof}: After the completion of this manuscript, 
we became aware of two recent publications 
by Guo and coworkers~\cite{Guo011,Guo012} which use a formalism based 
on NEGF and density functional theory similar to that described here. 
However, there is significant difference in detail between our approach 
and theirs.

\pagebreak

\begin{figure}
\centerline{\psfig{figure=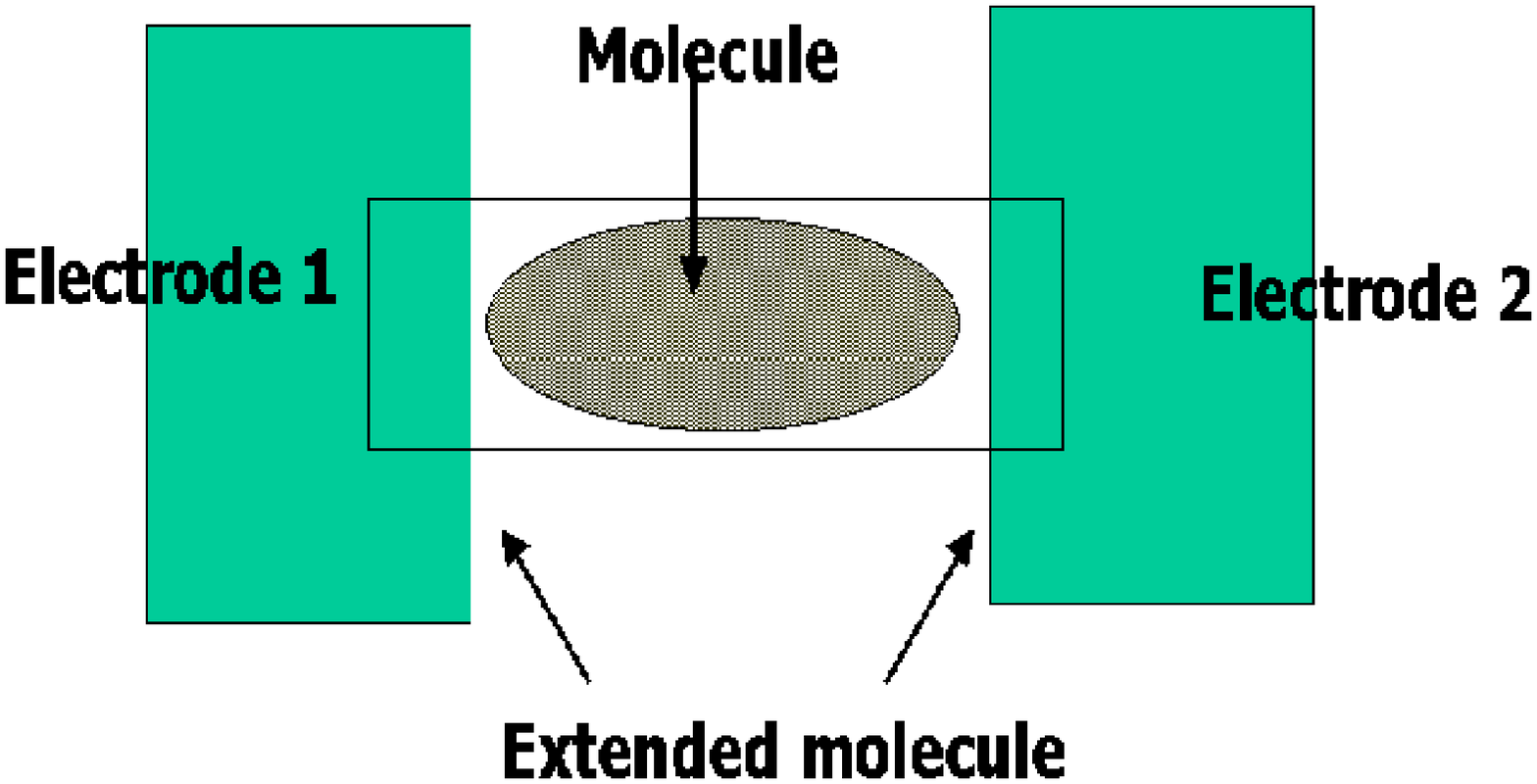,angle=0.,height=3.0in,width=4.0in}}
\vspace{1.0cm} 
\caption{Illustration of typical molecular devices.}
\label{Fig31}
\end{figure}

\begin{figure}
\centerline{\psfig{figure=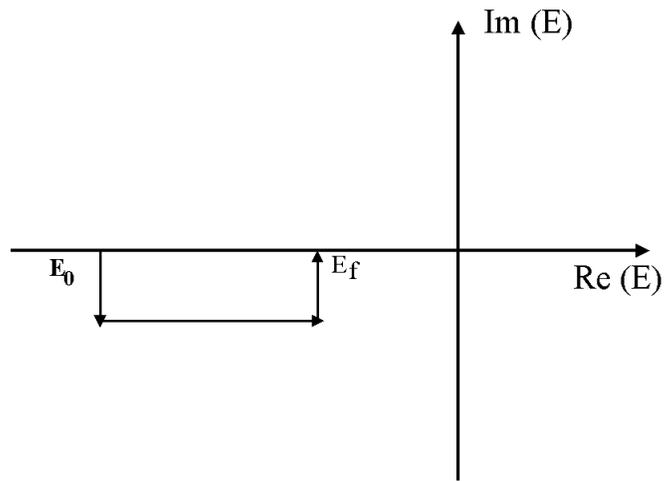,angle=0.,height=3.0in,width=4.0in}}
\vspace{1.0cm} 
\caption{Integration contour in the complex energy plane. The cutoff energy 
$E_{0}$ should be below the lowest occupied states of the system considered.}
\label{Fig32}
\end{figure}

\begin{figure}
\centerline{\psfig{figure=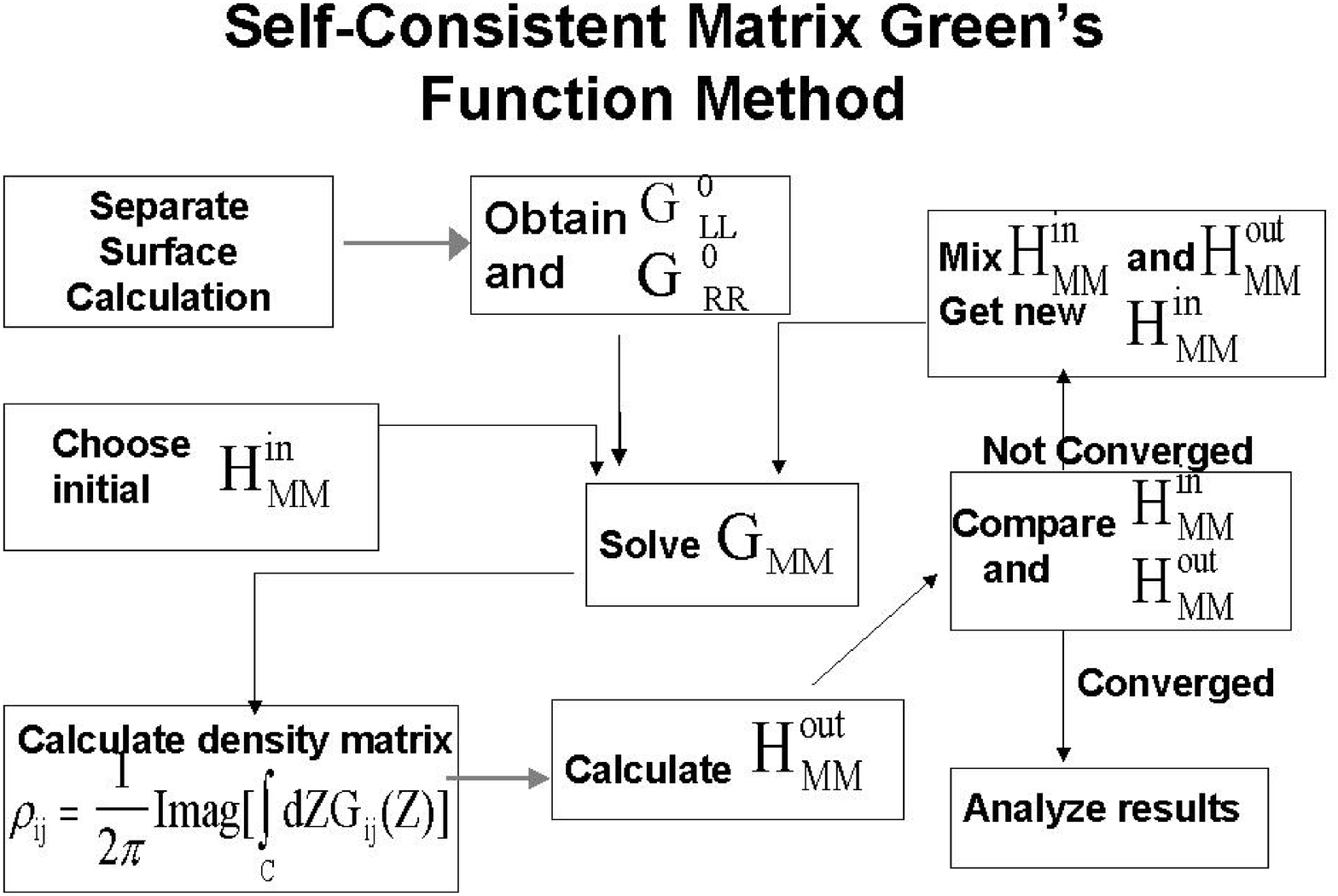,angle=0.,height=3.5in,width=5.0in}}
\vspace{1.0cm} 
\caption{Program flow chart of the self-consistent Matrix Green's function 
(MGF) approach to molecular device calculation. Note that besides the 
introduction of 
$V_{ext}$ and the self-energy operator $\Sigma_{L(R)}$, the only difference 
with the conventional molecular electronic structure calculation lies in the 
calculation of the density-matrix given the molecular Fock matrix. See the 
discussion in the main text. }   
\label{Fig33}
\end{figure} 

\pagebreak

\begin{figure}
\centerline{\psfig{figure=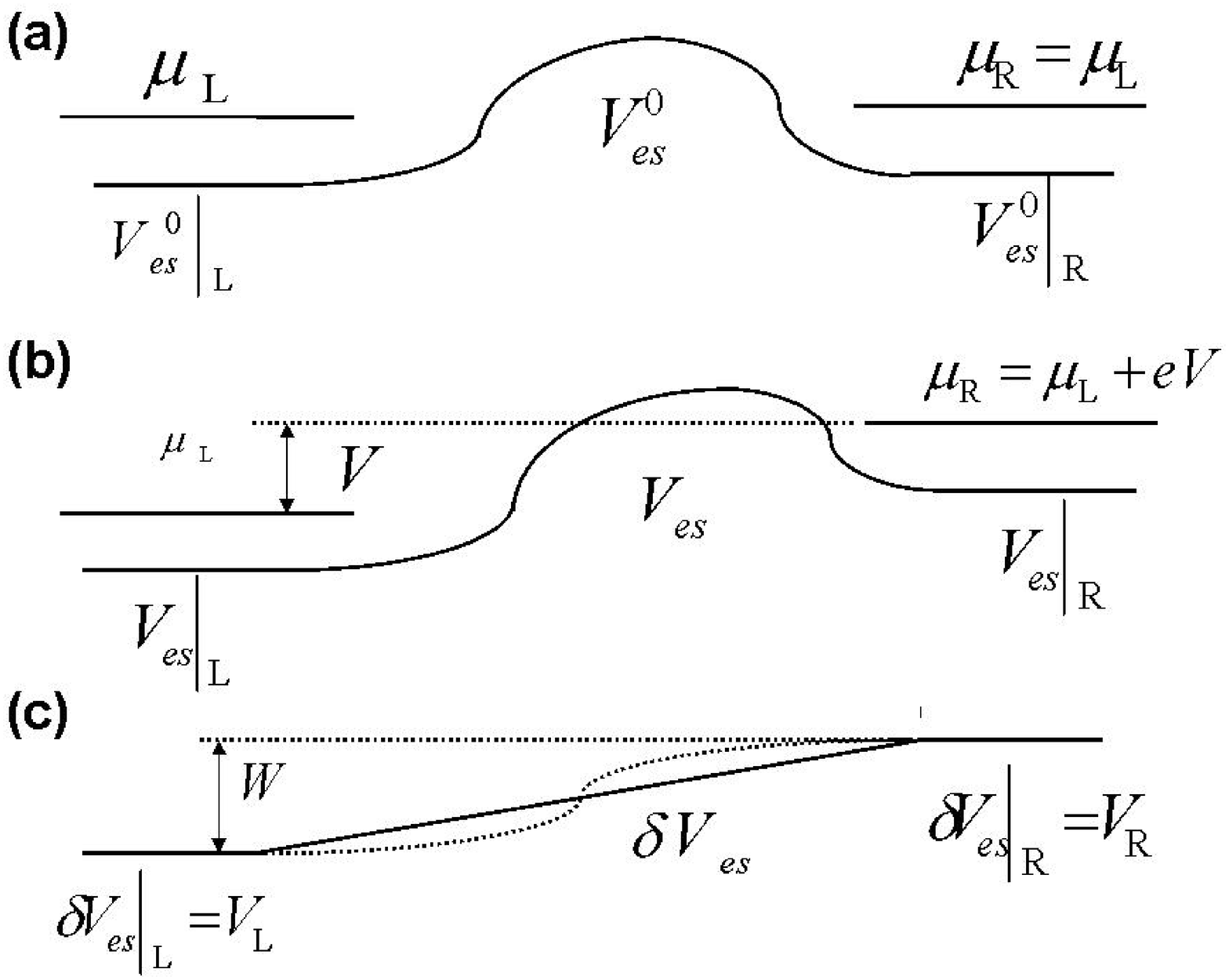,angle=0.,height=3.5in,width=5.0in}}
\vspace{1.0cm} 
\caption{(a) Schematic illustration of the electrostatic potential 
profile $V_{es}^{0}$ in the absence of applied bias. $V_{es}^{0}$ joins to 
the bulk values in the interior of the contacts. The exact value of 
$V_{es}^{0}$ should be obtained from the self-consistent calculation of 
the metal-molecule-metal junction as described in the main text. 
(b) Schematic illustration of the electrostatic potential profile $V_{es}$ 
in the presence of applied bias $V$. Note the changes in the boundary 
condition of $V_{es}$ under bias. (c) Difference in the electrostatic 
potential $\delta V_{es}=V_{es}-V_{es}^{0}$ in the presence and absence of 
applied bias $V$. Full line shows the linear ``applied bias'' $V_{bias}$ 
while dotted line shows the true $\delta V_{es}$ including the effect of 
the self-consistent charge redistribution $\delta \rho$ under bias (see 
Eq.\ (\ref{DelVes}) and discussions thererein). }
\label{Fig34}
\end{figure}

\end{document}